# UCLA Space-Time Area Law Model: A Persuasive Foundation for Hadronization

S. Abachi [a], C. Buchanan [b], A. Chien, S. Chun [c], B. Hartfiel [d]

Department of Physics and Astronomy, University of California, Los Angeles, CA 90095, USA

---

[a] email: sabachi@physics.ucla.edu

[b] email: buchanan@physics.ucla.edu

[c] Now at Boeing, 2260 East Imperial Hwy. El Segundo, CA 90245, USA

[d] Now at Physics Department, California State University Dominguez Hills, Carson, CA 90747, USA

**Abstract.** From the studies of rates and distributions of heavy quark (*c*, *b*) mesons we have developed additional evidence that hadron formation, at least in the simplest environment of $e^+e^-$ collisions, is dominantly controlled by a Space-Time Area Law ("STAL"), an approach suggested by both non-perturbative QCD and Relativistic String Models. From the dynamics of heavy quarks whose classical space-time world-lines deviate significantly from the light-cone, we report the exact calculation of the relevant space-time area and the derivation of a Lorentz invariant variable, $z_{eff}$, which reduces to the light-cone momentum fraction *z* for low mass quarks. Using $z_{eff}$ in the exponent of our fragmentation function in place of *z*, we find persuasive agreement with L = 0, 1 charmed and bottom meson data as well as for u, d, s L = 0 states. Presuming STAL to be a valid first-order description for all these meson data, we find the scale of other possible second order effects to be limited to ~20% or less of the observed rates. The model favors a *b*–quark mass of *~4.5* GeV.

# 1 Introduction

Hadronization – the evolution of a hard quark or gluon into a jet of hadrons – is a non-perturbative QCD process for which currently only phenomenological treatments are available. Figure 1 depicts the main three stages of hadron formation of light-quark mesons in the simplest environment of $e^+e^-$ collisions. First, the colliding beams annihilate and proceed through a virtual $\gamma$ or $Z^o$ which decays into the "primary" quark pair ($q_0, \bar{q}_0$) – a process calculable by electro-weak theory. The next step is the fairly well understood high $Q^2$ regime where these energetic quark and anti-quark fly apart. The behavior of the stretched color tube (string) between them and their subsequent hard gluon radiation is calculable to a fairly high level of accuracy by perturbative QCD and leading log techniques and have been implemented in several numerical simulations, most successfully by Webber [1] and Lund [2, 3]. We use the implementation in Lund's JETSET [2] program as input to our modeling of the next stage.

The last and, for us, most important major step of development (shown in Fig. 1 for light primary quarks) follows the process in which the primary quarks ($q_0, \bar{q}_0$) fly apart and stretch a narrow color field between them. The transition from this state to hadrons ("hadronization") is the focus of our study. By its nature, the soft process of hadronization, which develops as time evolves (upward on the page in Fig. 1), is not calculable via perturbative expansions because of the large value of the strong coupling at the very small momentum transfers in this stage. Therefore phenomenological models have been constructed to describe the process; most notably the cluster model of Webber [4] and the string model of Lund [5] – implemented by the Monte Carlo programs HERWIG [1] and JETSET [2, 3], respectively – as well as our own. Our implementation follows the Lund JETSET "outside-in iterative approach" in which the stretched color field between $q_0$ and $\bar{q}_0$ breaks up to produce new quark pairs such as $q_0\bar{q}_1$ and $q_1\bar{q}_o$, with each pair taking away a fraction of the energy; this implementation process is repeated for each new pair until the quarks in the pairs can be confined within ordinary colorless hadrons in a so-called yo-yo mode. The line segments in the upper part of Fig. 1 represent the final quark pairs, and the shaded squares show the produced hadrons (here only mesons are shown as di-quark pairs are needed to show baryons).

Historically, there are two distinct important roles which these models can play, with different objectives and criteria for evaluating their efficacy. One role is to give as accurate a description of the relevant data as possible, using as many parameters as are needed and where each parameter preferably has some plausible physical basis. Such models are useful, e.g. for, detector builders who need to design devices with a particular response or data analyzers who must know the acceptance of their detector to calculate particle rates, distributions, and correlations, etc. Programs of this variety, such as Lund's JETSET, have been quite successful. However, the many parameters involved tend to obscure the question of whether their physical bases are close to the mechanisms actually controlling the process. The center-piece of the Lund model is the derivation of the well known Lund Symmetric Fragmentation Function ("LSFF") [5], which is derived for massless (light) quarks in 1+1 dimensions and is given by

$$f(z) = N \frac{(1-z)^a}{z} e^{-b\frac{m_h^2}{z}} \qquad (1)$$

This function describes the probability density for producing a hadron with mass $m_h$ taking a fraction $z$ of the light-cone momentum ($p^+ = E + p$), where $z$ is defined below and $a$ & $b$ are arbitrary parameters arising naturally from the Lund approach. However, this function appears not to be appropriate for heavy quarks and therefore other fragmentation functions are used in the Lund implementation to describe them.

The other role, typified by UCLA's modeling, has the goal of making a persuasive case of identifying a dominant physical principle which controls the process. The measure of success in this approach is a combination of (a) the simplicity and attractiveness of the presumed underlying physical principle, (b) the smallness of the arbitrary parameter space, and (c) the quality of the agreement with the data.

The central thesis of our model is that of a Space-Time Area Law ("STAL") approach – suggested by both soft strong-coupled QCD [6, 7] and relativistic string models [8, 9] via a Least Action Principle – as the single dominant physical principle controlling the hadronization process. That is, whereas there may be other physical mechanisms involved (e.g., such as the tunneling-motivated s/u or wavefunction-motivated vector/all in Lund's model), our study suggests that they are at most secondary phenomena which would create relatively small corrections ≤ 20% to the rates predicted for various flavored hadrons by our STAL-based model.

Our earlier results [10] showed that the STAL approach worked persuasively for light quark mesons containing $u$, $d$, and s quarks. In this paper we extend our work to rates and energy/momentum distributions of charm and bottom mesons, including $L=1$ states. Again, the comparisons are rather persuasive for the proposition that the effects of STAL dominate the results.

## 2 UCLA scheme for light and heavy quarks

As described in detail in our earlier publication [10], the Space-Time Area Law (STAL) approach simply means that the probability of occurrence of an event is proportional to the negative exponential of the area in space-time swept out by the event – that is, $exp(-b'A_{plane})$, where an example of $A_{plane}$ is the 1+1 dimensional area shown in Fig. 1 for light quarks.

### 2.1 Light quark treatment

We have discussed [10] how the STAL assumption combined with the conservation of energy-momentum led to an event weight function, and using an iterative procedure – where one hadron at a time is pealed off from the end of an event – we arrived at the following fragmentation function [10] for light primary quarks in 1+1 dimensions:

$$f(z) = \frac{NC^2}{(4\pi)^2} \frac{(1-z)^a}{z} \left(1 - \frac{m_h^2}{Sz}\right)^a e^{-b'a_{xt}} \tag{2}$$

where $a_{xt}$ is the hatched area in Fig. 1 formed by the segments of the $q_o$ world-line and the $\bar{q}_1$ world-line and its extension. It can be shown that $a_{xt} = m_h^2/\kappa^2 z$ for light primary quark pairs (where $\kappa$ is the string tension of the order ~1 GeV/fm), so that

$$f(z) = \frac{NC^2}{(4\pi)^2} \frac{(1-z)^a}{z} \left(1 - \frac{m_h^2}{Sz}\right)^a e^{-b\frac{m_h^2}{z}} \tag{3}$$

where $C$ is the Clebsch-Gordan coefficient to combine the flavor and spin of the quark and anti-quark into the hadron; $N$ is a spatial "Knitting Factor" ~(2.7 fm)$^2$, presumed approximately the same for all hadrons, to knit the quark and antiquark into the hadron's spatial wave function; $S$ is $E_{c.m}^2$; $z$ is the light-cone momentum fraction defined for all quark masses by

$$z \equiv \frac{p^+_{hadron}}{p^+_{quark}} \tag{4}$$

(with $p^+ = E + p$). Expressions (2) & (3) are our UCLA Fragmentation Function "UCFF" for light quarks. Expression (3) is in the same form as the LSFF given by (1), with the exceptions of the small correction factor, $(1-m_h^2/Sz)$, and the important absolute normalization for any flavor-spin combinations given by $NC^2/(4\pi)^2$.

## 2.2 Heavy quark treatment

Figure 2(a) is the heavy quark analog of Fig. 1, showing the situation in 1 + 1 dimensions for a heavy primary quark pair. The curves depict (as will become apparent) the classical hyperbolic space-time curves (world-lines) of these quarks. For comparison, the dotted line segments are the light-cone paths for the case of a massless primary quark pair. The classical curved world-lines of heavy quarks naturally affect the areas spanned in certain segments of the event, in particular that of $A_{xt}$ – the hatched area in Fig. 2(a) bounded by the curved heavy-quark path and the $\bar{q}_1$ world-line and its extension. $A_{xt}$ so defined, is the analog of $a_{xt}$ in Fig. 1.

If one takes seriously the STAL approach and the use within it of classical motions for heavy primary quarks in a linear potential, then the light quark UCFF of (2) must be modified so that $a_{xt}$ is replaced by $A_{xt}$, and as a consequence UCFF of (3) must be modified such that $z$ is replaced by a new variable "$z_{eff}$" in the exponent. There have been approximate calculations of $A_{xt}$ and thereby of $z_{eff}$ [10, 11]. Since the values of $A_{xt}$ and/or of $z_{eff}$ are very important to a proper STAL treatment and they are somewhat sensitive to small approximations, an exact calculation of $A_{xt}$ has been performed by one of us[1] – but due to the rather long derivation the details will be published elsewhere. Here we sketch some useful steps in the derivation and present the final expressions for $A_{xt}$ and $z_{eff}$.

Consider the relativistic string model [5, 8, 9] of a bound $q\bar{q}$ pair (yoyo system). The statements that a linear potential exist, leading to a constant tension $\kappa$ which in turn leads to the equation of motion for each quark in the pair, are

$$\frac{dE}{dx} = \pm\kappa \qquad \frac{dp}{dt} = \pm\kappa \qquad (5)$$

It is generally favorable to work with light-cone variables in $xt$ and momentum spaces, defined by

$$x^{\pm} = x \pm t \qquad p^{\pm} = E \pm p \qquad (6)$$

---

[1] S. Abachi

Starting with (5) and using these variables, we can show that the space-time and momentum-energy areas as well as the light-cone momentum fraction variable $z$ are Lorentz invariant.

For heavy primary quarks the area of interest is $A_{xt}$ – the hatched area of Fig. 2(a). To calculate this, we begin with the equation of motion of the primary quark $q_o$ as given in (5). Since the areas are Lorentz invariant, for convenience we integrate the equation of motion in a boosted frame where $q_o$ is initially at rest as displayed in Fig. 2(b). Shown in the same figure, once the primary color tube breaks, a new quark-antiquark pair, $q_1 \bar{q}_1$, is created at the vertex point V, generating the first primary meson containing $\bar{q}_1$ and $q_o$. These quarks are accelerated towards each other until they cross at point E, where the primary meson is first created. By then, $\kappa \Delta x$ field energy is transferred to kinetic energy of the quarks and $\kappa \Delta t$ momentum is acquired by them. By integrating (5) for $q_o$, one obtains

$$\left(x - \frac{\mu}{\kappa}\right)^2 - t^2 = \frac{\mu^2}{\kappa^2} \tag{7}$$

This is a hyperbola describing the world-line of $q_o$ in the boosted frame and passing through $x = t = 0$. We can describe the motion of $\bar{q}_o$ in this frame as well by integrating (5) to get a different hyperbola. The lower hatched region in Fig. 2(b) represents $A_{xt}$. To evaluate this area, an integration between the hyperbolae curves from O to E is performed. This yields the following exact expression for the invariant $A_{xt}$

$$A_{xt} = \frac{1}{2\kappa^2}\left[\frac{m_h^2}{z} - \mu^2 - \mu^2 \log_e\left(\frac{m_h^2}{\mu^2 z}\right)\right] +$$

$$\frac{1}{2\kappa^2}\left[\varepsilon_p \mu + \mu^2 \log_e\left(1 - \frac{\varepsilon_p}{\mu}\right)\right] \tag{8}$$

where the second line term is small, since $\varepsilon_p$ is a small factor given by

$$\varepsilon_p = \frac{S}{2\mu}\left(1 - \frac{2\mu^2}{S} - \sqrt{1 - \frac{4\mu^2}{S}}\right)\left(\frac{m_h^2}{\mu^2 z} - 1\right) \tag{9}$$

The dominant first line term of expression (8) agrees with that of Bowler [11], and the second line term is an "additional term" (*A.T.*) which makes our result exact. Although the percent magnitude of *A.T.* for B-mesons produced at 90 GeV is negligibly small, it rises to ~10% for *c.m.* energies near the threshold for B-meson production. (Note that the *log* term in the *A.T.* is negative and relatively dominant and therefore *A.T.* is always negative.)

This completes the discussion of the Lorentz invariant area $A_{xt}$ which is used in

$$f(z) = \frac{NC^2}{(4\pi)^2} \frac{(1-z)^a}{z} \left(1 - \frac{m_h^2}{Sz}\right)^a e^{-b'A_{xt}} \tag{10}$$

This expression – the UCLA fragmentation function "UCFF" for all hadrons – is the analog of (2), where $a_{xt}$ is replaced by $A_{xt}$ in the exponent.

Based on $A_{xt}$, we can now define a new variable ($z_{eff}$) as an analog to the *z* variable, which leads to a heavy quark analog of UCFF of (3). Expression (3) can be recast from (10) for all quark masses as

$$f(z) = \frac{NC^2}{(4\pi)^2} \frac{(1-z)^a}{z} \left(1 - \frac{m_h^2}{Sz}\right)^a e^{-b\frac{m_h^2}{z_{eff}}} \tag{11}$$

comparing the exponent with that of (10) gives $z_{eff} = m_h^2/2\kappa^2 A_{xt}$ (with $b=b'/2\kappa^2$), which after replacing for $A_{xt}$ from (8) yields

$$z_{eff} = \frac{z}{1 - \frac{\mu^2 z}{m_h^2} - \frac{\mu^2 z}{m_h^2} \log_e\left(\frac{m_h^2}{\mu^2 z}\right) + \frac{2\kappa^2 z}{m_h^2} \times A.T.} \tag{12}$$

At the limit of $\mu \to 0$ (as for light quarks), $\varepsilon_p \to 0$, and thus $A.T. \to 0$. At this limit $z_{eff} \to z$, causing the UCFF of (11) exponent to be the same as that of LSFF's in (1). Furthermore, since the factor $(1-m_h^2/Sz)$ in (11) is very nearly unity, one may conclude that LSFF is a special case of our more general UCFF function, differing mainly by the presence of a modified variable $z_{eff}$ (instead of *z*) in the exponent. While for light mesons $z_{eff} \approx z$, for B-mesons $<z_{eff}> \approx 5<z>$, almost independent of *S*. The rates and distributions for B-mesons differ significantly from data if one uses *z* rather than $z_{eff}$ (equivalent to setting $\mu = 0$ in $z_{eff}$). We will show that the $z_{eff}$ expression, derived from STAL for hyperbolic quark world-lines in 1+1 dimensions, leads to satisfactory predictions of rates and energy distributions of heavy mesons in a natural way without having to interject any *ad hoc* procedure or parameters. This is unlike other current models where light and heavy quarks are treated separately.

We note that the UCFF as derived in (10) and (11) has the role of summarizing the consequences of our STAL assumption: (a) It replaces the z variable in the exponent of LSFF and converts it to a fragmentation function (UCFF) that is valid for heavy as well as for light hadrons with no additional free parameter. (b) The suppression of the production rates of heavier hadrons arises naturally from the hadron mass in the exponential and from the spatial/spin/flavor normalization $NC^2/(4\pi)^2$ for each hadron with no additional free parameters (whereas, for example, the production rate suppressions in the current Lund model are not due to hadrons mass, but occur via several presumed effects depending on variables such as secondary quark or di-quark masses or on whether the particle is a vector or of other types, with several adjustable parameters to control these effects.)

## 3 Analysis

Our methods of analysis are explained here and applied to light mesons (containing *u*, *d*, and *s* quarks), charmed mesons, and bottom mesons. For each sector, we examine the extent to which the STAL-based assumption holds.

### 3.1 Fitting and comparing with data

As described in [10], our six significant parameters are $\Lambda$ and $Q_o$ which control the parton shower, *a* and *b* in the fragmentation function, *n* which controls $p_t$ distributions, and $\eta$ which controls the suppression of multiple meson structures between a baryon-antibaryon pair. Recent data and our exact expression for $z_{\mathit{eff}}$ given by (12) have led to a slight retuning of the UCFF parameters *a* and *b* in our overall comparisons to the data. Compared to the values given in our earlier publication [10], the value of *a* was modified to 1.75 from 1.65, while *b* was changed to 1.10 from 1.18 (see [10] for a detailed description of the tuning process). Data, for this purpose, are mostly updated from the Particle Data Group ("PDG") tables [12], except for orbitally excited $D_s^{**}$ states at 91 GeV [13] and 10 GeV [14]. Estimated uncertainties in the decay branching fractions from higher mass states have been introduced in quadrature into the uncertainties for the data rates.

In Figs. 3(a–c) we compare available data [12–14] with the predicted rates for various flavored mesons at $E_{\mathrm{c.m.}}$ of 91, 29, and 10 (continuum) GeV. (Model predictions are shown even if usable data is not yet available.) One notes at each energy for each of the light quark, charmed quark, and bottom quark sectors that generally the production rates drop as the mass of the state increases and that our model predictions track the data rates rather well. Remarkably, this agreement involves data that span a factor of ~600 (~17 to 0.03 per event) in rates, a factor of ~40 (0.135 to 5.5 GeV) in hadron mass, a factor ~9 (10 to 91 GeV) in c.m. energy, all five accessible quark flavors and several spin states, including orbitally excited charmed and bottom mesons.

### 3.2 Criterion for a satisfactory agreement

Since our central thesis is that all physical mechanisms (e.g., s/u, V/all, etc) other than STAL control $\leq$ 20% of the various observed rates, a prediction deviating by more than 20% from data may signal a competing mechanism for STAL, unless it is due to statistical fluctuations. However if this difference measures more than two standard deviations, statistical fluctuations are the unlikely cause. Based on this argument we build an important analysis procedure; i.e., we interpret a prediction which is within 20% or two standard deviations of the datum as representing a satisfactory agreement, whereas a prediction which deviates from the datum by more than 20% and 2.0 standard deviations signals a potentially interesting effect. The latter is an indication of other processes that are large and potentially violate our assumption that STAL is the dominant underlying mechanism. The former is the criterion for confirming that STAL is in fact the dominant effect that controls hadronization and that any deviations are most likely statistical fluctuations. In the following three sections we apply this important criterion and examine each of the quark sectors in depth for any violations from our STAL assumption.

### 3.3 Light quark sector

Fractional rate deviations ($\delta$), defined by *(Mc–Data)/Data*, are calculated for all mesons at the three *c.m.* energies of 91, 29, and 10 GeV. Since there are no apparent significant dependences of these deviations on *c.m* energy (see Fig. 3), these three results are further combined by the usual weighted average method and the outcome is used throughout our analysis via our "criteria for satisfactory agreements".

For light mesons, in Fig. 4, we utilize the criterion set out in section 3.2 for a satisfactory agreement. In this two-dimensional figure we show the fractional deviations ($\delta$), versus the number of standard deviations ($n_s$) that the model differs from data for: (a) *u* & *d* and (b) for light mesons, containing *s*-quarks using the data combined from $E_{c.m.}$ = 91, 29, 10 GeV. The shaded areas in Fig. 4 represent regions which violate our criterion for a satisfactory agreement. An entry in these regions signals a deviation from the datum by more than 20% and 2.0 standard deviations. Two of the 11 entries (φ and η) are at the edge of shaded regions; the rest of the points are well within the satisfactory domain. At the three energies, φ is consistently over-predicted by of ~25–30% (~2σ), and the η is under-predicted by ~20% (3.7σ). The data rates for the displayed mesons range from 0.044 for φ, to 17 for $\pi^{\pm}$ – a range factor of ~400.

Representative fractional momentum ($x_p$ = $p_{hadron}/p_{beam}$) distributions from ALEPH [15] for $\pi^{\pm}$, $K^o$, $K^{*o}$, $\rho^o$, $\eta'$, φ, and η are compared with our model in Figs. 5(a–c). Though not perfect, the agreements are adequately good for our STAL-based assumption. Although not shown here, we arrive at similar conclusions for data distributions at 10 and 29 GeV. (See our earlier publication [10] for a more complete display of flavored meson momentum distributions and also of topological event distributions.)

From the above comparisons, we find the available light quark data in accord with the hypothesis that STAL is the dominant underlying physical principle, though there may be second order effects ~20%.

Having reaffirmed our earlier success [10] concerning the light quark sector with recent data and the slightly retuned values of *a* and *b*, we next turn to mesons containing heavy quarks. We use only STAL with the $z_{eff}$ but no new parameters other than reasonable $m_b$ and $m_c$ values, and equip our model with the orbitally excited (*L=1*) states of D and B-mesons, often referred to as $D^{**}$ and $B^{**}$ states.

### 3.4 Charm sector

As discussed, in Figs. 3(a–c) the model predictions follow the charmed meson data well at three different energies, extending the range of the model down another order of magnitude and encompassing massive orbitally excited states. Further analysis is introduced by our criterion for a satisfactory agreement through the two dimensional plot of δ versus $n_s$. This is shown in Fig. 6, which shows satisfactory agreements for all charmed mesons with only $D^{\pm}$ at the boundary of a shaded region. The data rates for the displayed mesons range from 0.004 for $D_s^{**}$, to 0.917 for $D^{*\pm}$ – a range factor of ~230.

The individual rates at the three *c.m.* energies do suggest that in the pseudo-scalar subsector we might be over-predicting non-strange charmed meson rates by ~20% and under-predicting strange charmed meson rates by ~30%. However, at 91 and 29 GeV, these predictions are strongly influenced by the B-meson decay branching fractions, which are not very well known and could be outside our estimated uncertainties.

More precise $D^*$ data have recently become available, and in Fig. 7 we compare our predicted $x_E$ spectrum for the $D^{*\pm}$ mesons with data from ALEPH [16] at 91 GeV. Contributions from $Z \rightarrow c$ dominates in the $x_E > 0.3$ region while $Z \rightarrow b \rightarrow c$ and gluon splitting ($g \rightarrow c\bar{c}$) contribute at low $x_E$. The dotted curve shows our predictions for $m_c = m_b = 0$ (i.e. $z_{eff} \rightarrow z$, with no gluon splitting but with *b*-enriched channel) which, as expected, deviates rather significantly from data at high $x_E$. With $m_c=1.25$ GeV, we predict the dashed curve which is consistent with high $x_E$ data but deviates from data at low $x_E$. This deviation is consistent with the lack of gluon splitting in our model. The solid curve is our prediction including the hard gluon splitting contribution as determined by the ALEPH collaboration [16, 17]; this agrees well with data at all $x_E$. The average $x_E$ value from this curve (0.387) compares quite well with the ALEPH datum [16] of 0.391±007.

ALEPH has separately calculated the rates in the hadronic, *b*-enriched ($Z \rightarrow b \rightarrow c$), and *c*-enriched ($Z \rightarrow c$) channels for some of the charmed mesons, including $D^*$, $D_s^*$ [16], and orbitally excited narrow charged states of $D_{s1}^*$ and $D_{s2}$ [13]. The rates, in terms of the probability that these mesons materialize in each channel × 100, are compared with our predictions in Table 1. In the *c*-enriched channel we have satisfactory agreements for all cases although the data relative uncertainties are very low (only few % higher than those of the total rates). In all other cases we have satisfactory agreements as well. Since data in Table 1 are all within two standard deviations from model predictions, they all comply with our criterion for a satisfactory agreement and no further examination of them is required. In the *c*-enriched sample ALEPH measures [16] an average scaled energy of $\langle x_E \rangle_c = 0.4878 \pm 0.0076$ for $D^{*\pm}$, which is 1.5$\sigma$ lower than our prediction of 0.4993.

Based on the above comparisons, the available charmed meson data are compatible with the assumption that STAL is the dominant underlying principle for hadronization.

**3.5 Bottom sector**

Ideally, the success of the STAL-based model using $z_{eff}$ is best tested with B-mesons since the quark mass effect contribution to the $z_{eff}$ variable is maximal. Table 2 offers a comparison of the available B-meson data [12] with our model. As expected, the agreements are poor with $m_b=0$ (equivalent to using $z$ rather than $z_{eff}$), especially for the complicated overlapping $L=1$ $B^{**}$ states which is under-predicted by 4.7 standard deviations and ~97%. Using $m_b = 4.5$ GeV in $z_{eff}$, the $B^{**}$ rate improves to 1.8 standard deviations and 36% below the data.

In Fig. 8, we examine the data further through our criterion for a satisfactory agreement. This shows that all available B-mesons are in the satisfactory region. In addition, Table 2 indicates good agreement between data and the model for B, $B_s$, and $B^*$, and fairly good for $B^{**}$ mesons.

ALEPH [18] and SLD [19] have measured the $x_E$ spectra of inclusive weakly decaying $B^{\pm}$ hadrons. These data are compared in Fig. 9(a) with our predictions. Unfortunately, here our model depends on the decays of higher mass hadrons with poorly known masses and branching fractions. Shown in Fig. 9(b), ALEPH [18] has also given an $x_E$ distribution for "primary" $B^{\pm}$ hadrons, i.e. those that are not decay products of higher mass states. In this case, our model prediction is less ambiguous, but potentially uncertain model dependence is involved in the data analysis.

As expected, in both cases the predicted spectra for $m_b=0$ (dashed curves) are much harder than the data. The $m_b=4.5$ GeV mass value gives much softer spectra (solid curves) that come close to agreement with data but do not quite reproduce the shape in detail. For Fig. 9(a) (the spectra of inclusive weakly decaying $B^{\pm}$ mesons), our model predicts an average $x_E$ of 0.712, about 0.5σ smaller than the ALEPH value of 0.716±0.008, and 0.6 σ larger than SLD value of 0.709±0.005.

Although, the *b*-sector potentially offers a very effective test of the STAL assumption, unfortunately the experimental situation is rather less favorable than for charm: data errors are generally large and usable rate data are only measured at one *c.m.* energy for two particles and two additional categories of $B^*$ and $B^{**}$ which are sums over higher spin states; comparisons of spectra, in one way or another, involve model dependencies. Furthermore, several inputs to the model such as *b*-baryon masses are unmeasured or poorly known. These experimental issues may explain the status of the $B^\pm$ spectra of Fig. 9. Given these considerations, the *b*-sector data currently seems to be consistent with the idea that STAL is the dominant underlying principle.

## 4 Discussions

Two important issues – the mass of the heavy quarks and possible other non-STAL based contributions – deserve further discussions. Below we study these issues.

### 4.1 Heavy quark masses

In the *b*-meson sector we have used a mass value of 4.5 GeV for the *b*-quark. This value is adopted because we observe that our model predictions for the rates and spectra of B-mesons are sensitive to the *b*-quark mass, and the model seems to favor this value when $m_b$ is varied. However, due to the experimental inadequacies surrounding the *b*-hadron data, this value may not be highly optimized.

We point out that the parameters of the model (*a* & *b*), which appear in our fragmentation function, are not strongly correlated with $m_b$. We observed negligible changes, if any, in the light meson rates by varying the *b*-quark mass within any reasonable range. This simplifies the following studies quite a bit, as only the spectra and relative rates of B-mesons are significantly affected by $m_b$. This lack of dependence occurs because the $Z \to b\bar{b}$ branching fraction is independent of the *b*-quark mass. We observed that when the mass is varied from 3.9 to 4.9 GeV, the $B^{**}$ rate prediction increased sharply, while the rate prediction declined somewhat sharply for $B^*$ and declined softly for B. Any change in the $B^{**}$ rate should be compensated by variations in the $B^*$ and/or B rates. Furthermore, for both the leading and weakly decaying $B^\pm$ meson (which we also studied in the *b*-sector) the mean $x_E$ of the spectra declined sharply with $m_b$. We are already familiar with the extreme cases in Figs. 9(a & b) where $m_b \to 0$ led to much harder energy spectra for $B^\pm$ production. In fact, due to the $Z \to b\bar{b}$ branching fraction constraint, the variations with $m_b$ are expected to be sharper for the mean fractional energies that they are for the rates.

For each mass, we compared the B-meson (B, B$^*$, B$^{**}$) rate predictions with data [12] and compared the mean fractional energy predictions ($<x_E>$) for weakly decaying and/or for leading B$^\pm$ meson with that of ALEPH [18] and SLD [19] data. Figure 10 summarizes all the comparisons by plotting the over-all $\chi^2 = \sum (Data - MC)^2 / \sigma^2$ versus several b-quark mass assumptions. The favored mass appears in the neighborhood of $m_b \approx 4.5$ GeV. In this figure we see a rapid rise in $\chi^2$ when one departs from the mass region of 4.1–4.7 GeV. Although further improvements in the b-hadron data are likely to lead to a more accurate fit for $m_b$, the result as it stands is impressive as it signals consistency between our model predictions and the established range of acceptable b-quark mass.

Our predictions are not very sensitive to the charm quark mass. Although for the extreme assumptions that $m_c = m_b = 0$ (i.e. $z_{eff} \to z$) the model distribution in Fig. 7 deviates somewhat significantly from data, reasonable c-quark mass variations do not seem to alter the rates or $x_E$ spectra noticeably. We thus adopt the common value of 1.25 GeV for the c-quark mass.

### 4.2 Possible other significant factors and higher order effects

In addition to our STAL-based modeling – which appears to control most of the observed production rates for the various flavored mesons – there are other possible physical mechanisms which could influence production rates and spectra such as: suppression parameters (e.g. s/u or vector/all which are employed in the Lund model), incorporating scalar and tensor states, better b-hadron production data, and inclusion of secondary heavy quark production.

There are cases of particular interest:

1. If one examines the model-data comparisons for any indication of a vector/all type of suppression, one finds that there is no need for such suppression. That is, in our model, which uses masses via STAL and Clebsch-Gordon coefficients to control production rates, there is no need to use any factor for the ratio of vector to pseudoscalar production other than the natural 3/1 ratio arising from the counting of final states.

2. The situation for strange versus non-strange production is more interesting:

a) The pseudoscalar η is under-predicted by ~20% (3.7 standard deviations). It is also suggestive that the η′ might be under-predicted by a similar percentage.

b) The vector φ is over-predicted by ~25-30% (~2.1 sigma).

c) The pseudoscalar non-strange $D^{\pm}$ and $D^o$ are over-predicted by ~20% (~2.0 sigma) whereas the $D_s$ is under-predicted by ~30% (1.6 sigma).

3. The situation for the $L=1$ $B^{**}$ set of states in the $b$-sector also requires attention. Our model under-predicted the $B^{**}$ rate by ~1.8 standard deviations and ~36% (but it passes our criterion for a satisfactory agreement). The predicted energy spectra for the charged B-meson follow the data moderately well, but some discrepancy is observed. In the $b$-sector we argued – based on several experimental inadequacies – that the source of the effect may be data related.

4. Due to their very small rates, our approach does not yet model any secondary $c$ and $b$ quarks production process. The ALEPH studies [16, 17] utilized in section 3.4, show that at 91 GeV only ~3% of the non-primary quarks are $c\bar{c}$ pairs from gluon splitting. We infer that contributions of secondary $b\bar{b}$ pairs at all three energies and of $c\bar{c}$ pairs at 29 and 10 GeV are negligible and that only small corrections from secondary charm production would be necessary at 91 GeV. In our following statistical analysis, this small effect is contributing a bit to the total of other possible non-STAL based phenomena.

## 5 Statistical analysis of STAL

In the previous section we saw some cases emerge as candidates for study for possible interesting higher order effects beyond STAL. Here we estimate the scale of the sum of all such possible effects as follows: For the 53 total production rate data points (in three energies) and with six significant parameters ($a$, $b$, two parameters controlling the parton shower, $P_t$, and baryon production), we have a chi-squared ($\chi^2$) of 79, for 53–6 = 47 degrees of freedom (*DOF*) – a $\chi^2$/*DOF* of ~1.7. We incrementally add additional Gaussian uncertainties in quadrature to each data point in our model until a $\chi^2$/*DOF* of 1.0 is obtained. This added uncertainty is ~20%. We interpret this 20% value as an estimate of the scale of the cumulative effect of all possible second order phenomena, if any.

## 6 Baryons

Similar to our conclusion in reference [10], Fig. 11 shows a reasonable match when ALEPH data spectra [15] for p and Λ are compared with predictions of our model. However, examining the available baryon data [12] versus our rate predictions, the entries in the two dimensional plot of Fig. 12 show only four (p, Λ, $\Delta^{++}$, $\Sigma^o$) of the nine baryon entries in the satisfactory region, whereas five ($\Sigma^\pm$, $\Xi^-$, $\Xi^{*o}$, $\Sigma^{*\pm}$, $\Omega^-$) violate our criterion for a satisfactory agreement. The current model [10], includes one free parameter to control multiple meson production between a baryon-antibaryon pair. However, baryons are 3-quark states requiring a more complicated mechanism than is shown in Fig. 1. Some of these complex issues, such as di-quark production and flavor chaining arrangements involving intermediate "popcorn" meson production between baryon and antibaryon, are considered and discussed in our earlier publication [10]. Given such complexities, our comparisons here are encouraging, but challenging. Precise data on baryon-antibaryon and baryon-meson correlations and direct evidence of multiple meson structures ("popcorn") between baryon and antibaryon pair would be very useful toward better understanding the mechanisms of baryon formation.

## 7 Conclusions

Existing hadronization models have placed emphasis on accurate reproduction of data by Monte Carlo programs. The output accuracy of such models is typically controlled by adjusting many parameters, each presumed to be justifiable by some physics explanation and having roughly comparable significance in terms of controlling the production rates. By contrast, our model has placed emphasis on identifying a single suitable underlying principle with a very small set of ~6 parameters. By adopting a QCD-inspired Space-Time Area Law (STAL) as our single underlying principle, we have been able to persuasively describe all available data on meson production rates and spectra in $e^+e^-$ collisions while limiting other possible second order effects to a scale of ~±20%. The agreements involve data that span factors of ~600 in rate (0.03–17 per event), ~40 in mass (0.135–5.7 GeV), ~9 in *c.m.* energy (10–91 GeV), and all five accessible quark flavors with several spin states, including orbitally excited charmed and bottom mesons.

Beginning with the STAL assumption and the dynamics of heavy quarks whose classical world-lines are hyperbolic, we arrive at a Fragmentation Function (UCFF) which is valid for heavy mesons as well as for light mesons, with no additional free parameter. The expression for UCFF has essentially the same functional form as that of the Lund Symmetric Fragmentation Function (LSFF), only its $z$ variable is replaced in the exponent by a new Lorentz invariant variable $z_{eff}$ which depends on the quark mass $\mu$ and reduces to the usual $z$ variable as $\mu \to 0$ for light quarks. We have reported an exact calculation of the relevant space-time area ($A_{xt}$) from which the exact $z_{eff}$ variable can be derived. By varying the $b$-quark mass, we found that our model favors a value of ~4.5 GeV, well within the known mass range for this quark. In our model, the relative production rates of states of different spin and flavor are controlled naturally by Clebsch-Gordan coefficients and the hadron mass dependence inherent in STAL.

Presuming STAL to be the dominant underlying physical mechanism, we have estimated the combined effects of possible second order phenomena, if any; e.g., strange quark (s/u) and vector suppressions (vector/all), etc. We conclude that the current state of data is consistent with no vector/all suppression. There are a few potentially interesting strange/non-strange comparisons involving $\eta$, $\eta'$, $\varphi$, $D^{o\pm}$, and $D_s$, as well as $B^{\pm}$ spectra. We are consistent with the idea, beginning with the STAL basis, that the assumptions of strangeness and/or vector suppressions are wholly or partially artificial. Further investigation on the sources of the differences requires additional data such as higher mass states that are not yet included (for example, scalar and tensor mesons) and other improvements such as STAL-based modeling of secondary heavy quark production.

There is more to be done on several fronts. Larger discrepancies were found in the baryon sector. More complex baryon models could be introduced, but they require additional data on baryon–antibaryon and baryon–meson correlations as well as direct evidence of intermediate meson structures ("popcorn") between baryon and antibaryon. The spectroscopy of excited bottom states must advance in order to develop this area. Our discrepancy in the inclusive B-meson spectrum may be due to bottom baryons and excited mesons. Accurate measurements of charmed and bottom baryons would be especially useful. We anticipate that the eventual resolution of the above issues through high statistics, high quality data such as at Babar and Belle will reveal further information about hadron formation dynamics.

*Acknowledgements.* We are grateful to the Lund group for the usage of their JETSET program within our model implementation. We also thank David Muller (SLAC) for many discussions about SLD data and for his comments. This work was supported in part by the US Department of Energy, under grant number DE-FG03-91ER40662.

**Captions**

**Fig. 1.** Hadronization in $e^+e^-$ collisions has three distinct regions: (I) Fully described by electro-weak theory, the $e^+$ and $e^-$ annihilate to a pair of massless quarks ($q_o, \bar{q}_o$), which begin to fly apart along their light-cones. (II) Described by perturbative QCD, in the region denoted by $A_{\text{plane}}$, the stretched color field between primary quarks radiates gluons that can further turn into other gluons and quark-antiquark and di-quark pairs (parton shower). (III) From breakings of the color fields, pairs such as $q_o\bar{q}_1$ and $q_1\bar{q}_o$ are produced in our implementation which subsequently can break into other quark-antiquark pairs. Finally a stage is reached (line segments in the upper part of figure) where the produced pairs can be confined inside colorless hadrons (shaded rectangles). The event space-time area marked by $A_{\text{plane}}$ (bordered by the solid line segments) and $a_{\text{xt}}$ (hatched area), are related to event and hadron formation probabilities, respectively, and the hadron box area $A_h^{xt} = m_h^2/2\kappa^2$.

**Fig. 2.** (a) The analog of Fig. 1 for a heavy primary quark pair is shown. While massless quarks travel on the light-cone, heavy primary quark classical world-lines in 1+1 dimensions are curved and they modify the space-time areas. $A_{\text{xt}}$ (hatched region) and $A_{\text{plane}}$ (the event area, bordered by the solid line segments and curves) are the modified areas. Therefore event and hadron formation probabilities are also modified. (b) The hyperbolic world-lines of the heavy quarks $q_o$ and $\bar{q}_o$ in a boosted frame where $q_o$ was initially at rest. The lower hatched area $A_{\text{xt}}$ (a Lorentz invariant) is shown, which is bounded by the $q_o$ and $\bar{q}_o$ curved paths and the light-cone path segment of $\bar{q}_1$ and its extension.

**Fig. 3.** All available data rates for meson production are compared with the UCLA model predictions at *c.m.* energies of (a) 91 GeV, (b) 29 GeV, and (c) 10 GeV, in terms of the number of mesons produced per event. Particle names and mass value labels for each entry are displayed in these plots. Model predictions are shown even if usable data are not available.

**Fig. 4.** Relative rate deviations (δ), defined by *(Mc–Data)/Data)*, versus number of standard deviations ($n_s$) that the model predictions deviate from data, are plotted for (a) light $u, d$ mesons, and (b) for light strangeness-containing mesons. Data are combined from 91, 29, and 10 GeV energies when available. An entry in a shaded region signifies a model prediction which deviates from the datum by more than 20% <u>and</u> 2.0 standard deviations. Such entries violate our "criterion for satisfactory agreement" and may signal a potentially interesting mechanism competing with STAL.

The data rates per hadronic event for the displayed mesons range from 0.044 for $\varphi$, to 17 for $\pi^\pm$ – a range factor of ~400.

**Fig. 5.** ALEPH $x_p$ ($p_{hadron}/p_{beam}$) spectra for various mesons at the Z-peak are compared with our STAL-based model. Symbols are data and solid curves are the model predictions.

**Fig. 6.** The analog of Fig. 4 ($\delta$ versus $n_s$) is shown for charmed mesons. Entries outside shaded regions signify satisfactory agreements. The data rates per hadronic event for the displayed mesons range from 0.004 for $D_s^{**}$, to 0.917 for $D^{*\pm}$ – a range factor of ~230.

**Fig. 7.** The ALEPH $D^{*\pm}$ $x_E$ spectrum at 91 GeV compared to our predicted $x_E$ spectrum. Contributions from $Z \to c$ dominate in the $x_E > 0.3$ region while $Z \to b \to c$ and gluon splitting ($g \to c\bar{c}$) contribute at low $x_E$. The dotted curve shows our predictions for $m_c = m_b = 0$ (i.e. $z_{eff} \to z$). Dashed curve is with $m_c = 1.25$ GeV. The solid dots show the gluon splitting contribution, extracted from ALEPH data [16, 17]. The solid curve is our final prediction with $m_c = 1.25$ GeV and with the gluon splitting contribution added.

**Fig. 8.** The analog of Fig. 4 ($\delta$ versus $n_s$) for B-mesons at 91 GeV. Entries outside shaded regions signify satisfactory agreements. The data rates per hadronic event for the displayed B-mesons range from 0.118 for $B^{**}$, to 0.288 for $B^*$.

**Fig. 9.** The $x_E$ spectra of the $B^\pm$ mesons (a) from ALEPH and SLD for weakly decaying and (b) from ALEPH for leading B hadrons are compared with UCLA predictions. Model results are given both for $m_b = 0$ (dashed curves) and $m_b = 4.5$ GeV (solid curves).

**Fig. 10.** The plot of $\chi^2 = \Sigma[(MC - Data)/Error]^2$ for several $b$-quark mass assumptions, where the terms in the sum are B, $B^*$, $B^{**}$ meson rates as well as mean $x_E$ values for leading and/or weakly decaying $B^\pm$ spectra of ALEPH and SLD. Based on this result, $m_b \sim 4.5$ GeV assumption has the minimum $\chi^2$, with a $\chi^2/DOF$ of 1.6.

**Fig. 11.** Proton and $\Lambda$ $x_p$ spectra from ALEPH are compared with the UCLA results. Symbols are data and solid curves are the model predictions.

**Fig. 12.** The analog of Fig. 4 (δ versus $n_s$) for baryon data combined from 91, 29, and 10 GeV where available. Entries in the shaded regions indicate lack of satisfactory agreements.

**Table 1.** Comparisons of the UCLA model predictions with the ALEPH results for several L = 0, 1 D-mesons in hadronic, *c*-enriched, and *b*-enriched events. The values are in terms of percent probability that these mesons materialize in each channel. For example, for D* in the *c*-enriched channel, this would be 100 × (# of D*'s in Z→*c* jets) / (# of Z→*c* jets). Last column shows $n_s$, the number of standard deviations that the model differs from data. Last row is the total $\chi^2 = \Sigma[(MC - Data)/Error]^2$ from all data in the table which is for 9 *DOF*.

**Table 2.** We compare the B-meson data rates [12] with the UCLA model for *b*-quark masses of zero and 4.5 GeV. In this table, $n_s$ is the number of standard deviations that the model differs from data, i.e., *(MC-Data)/Error*. The last row shows $\chi^2$'s (for 4 degrees of freedom) for $m_b=0$ and for $m_b=4.5$ GeV choices. B* & B** values are for all charge states, i.e., for *bu*, *bd*, and *bs* states combined.

**Fig. 1**

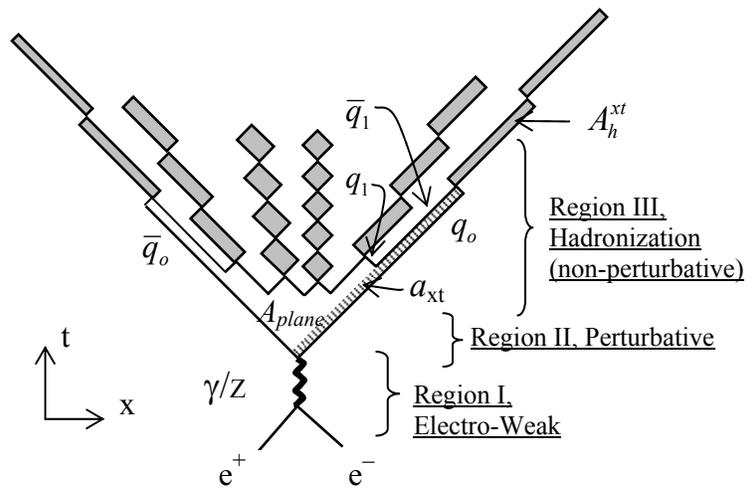

**Fig. 2**

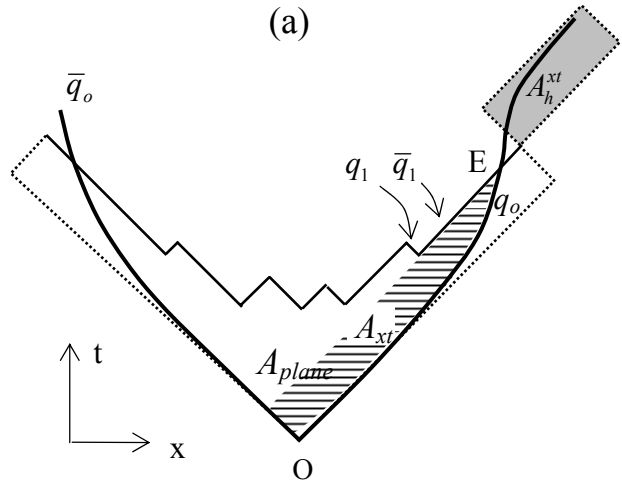

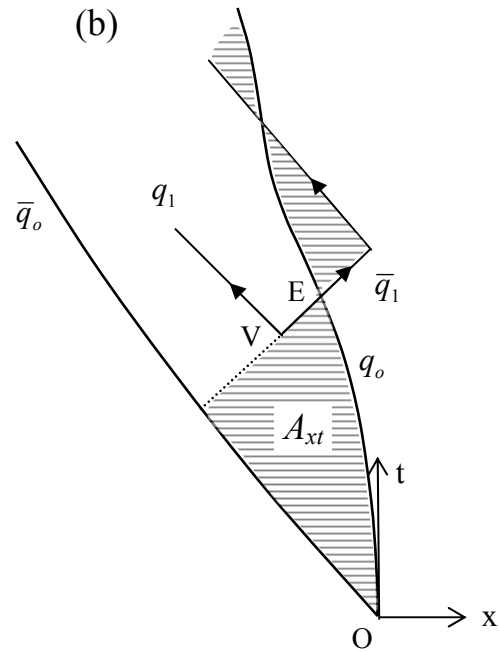

**Fig. 3**

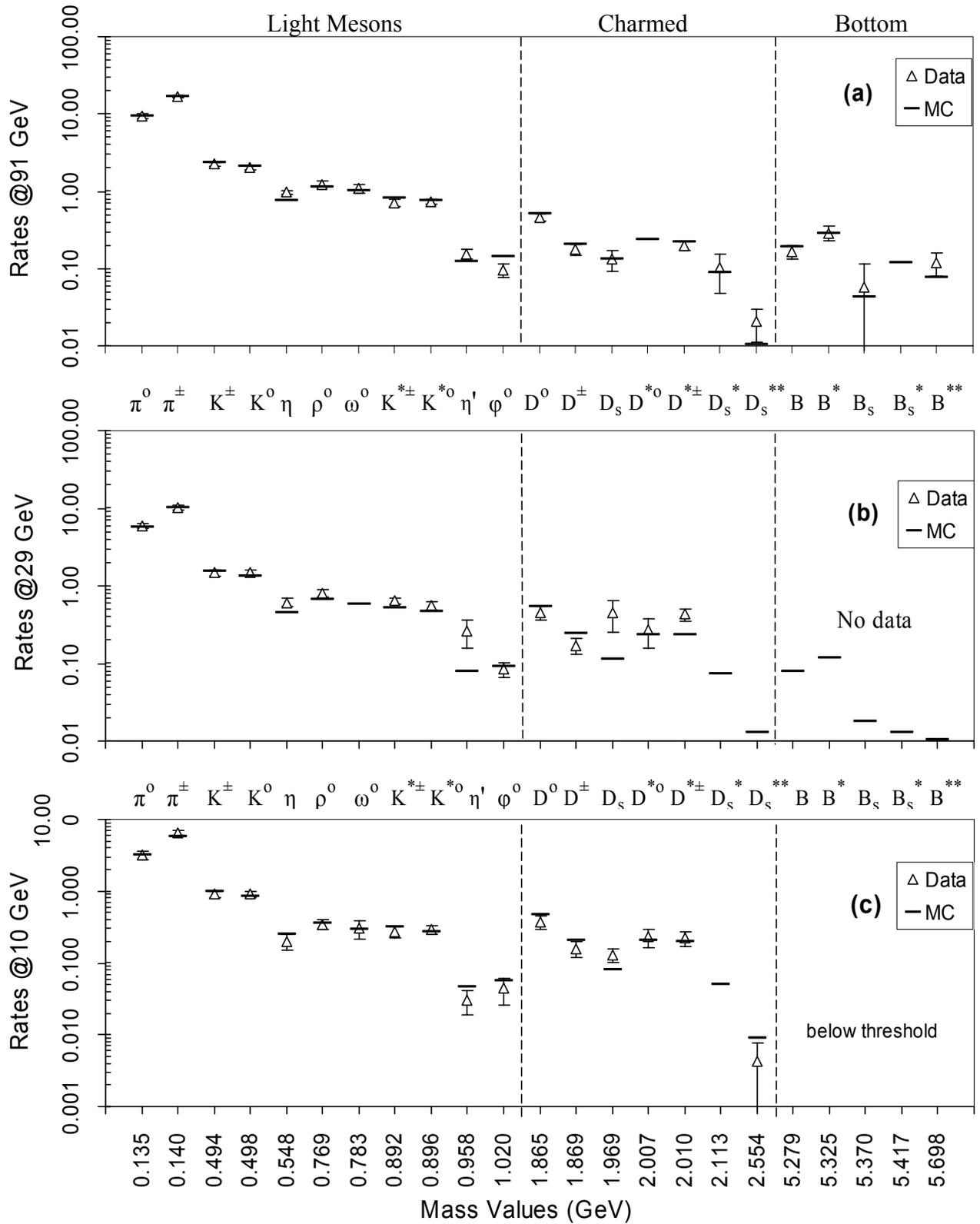

**Fig. 4**

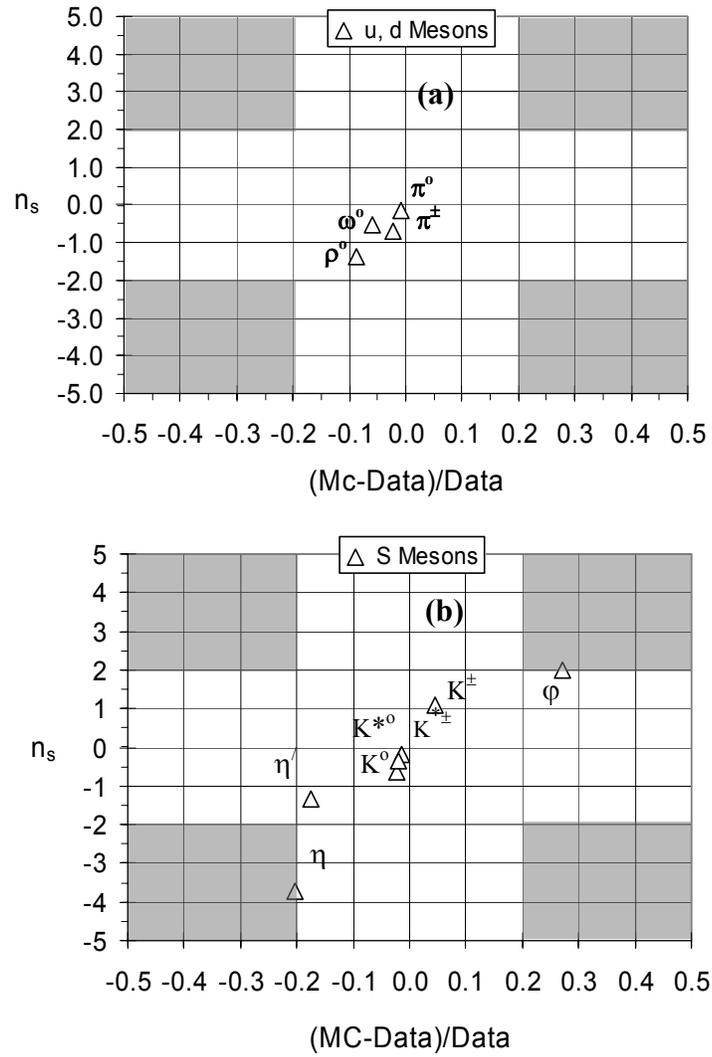

**Fig. 5**

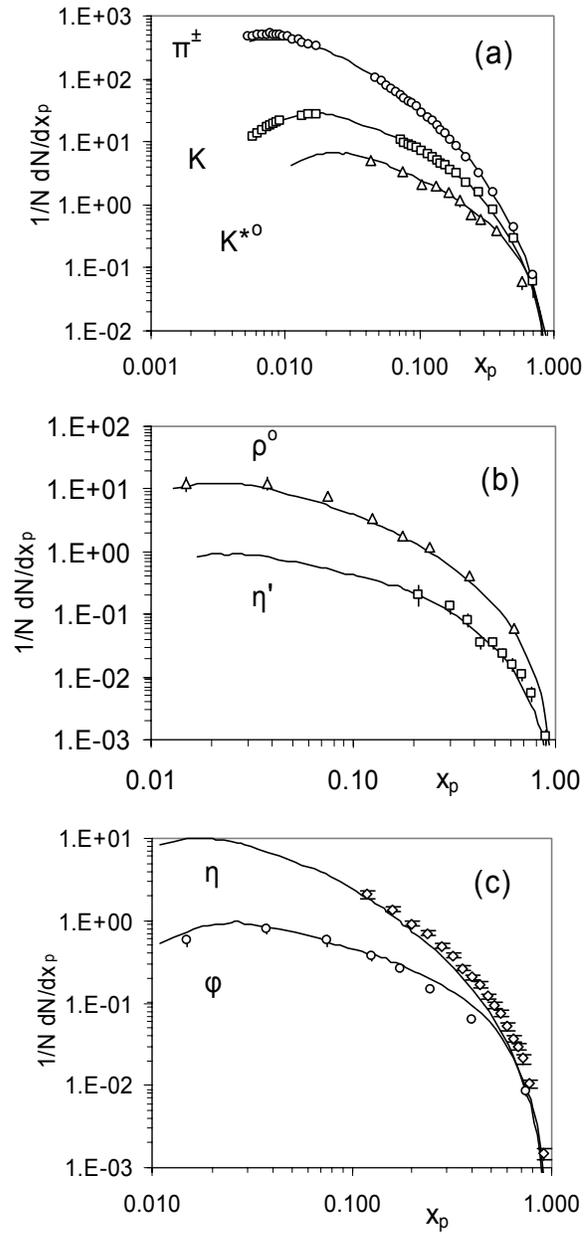

**Fig. 6**

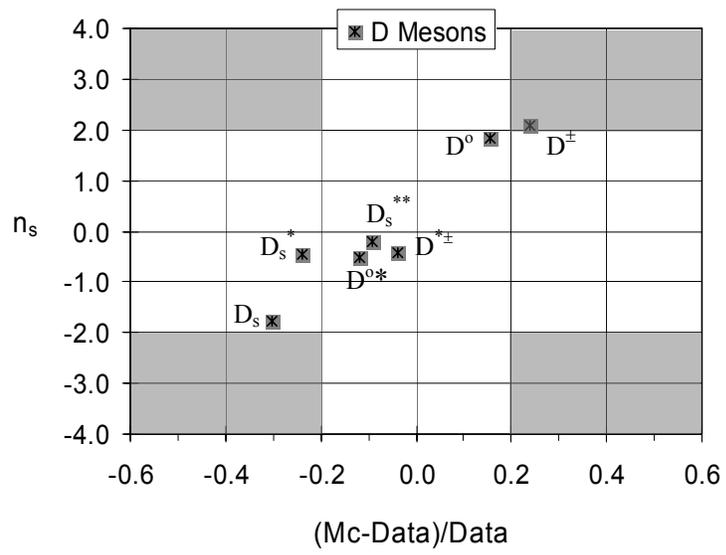

**Fig. 7**

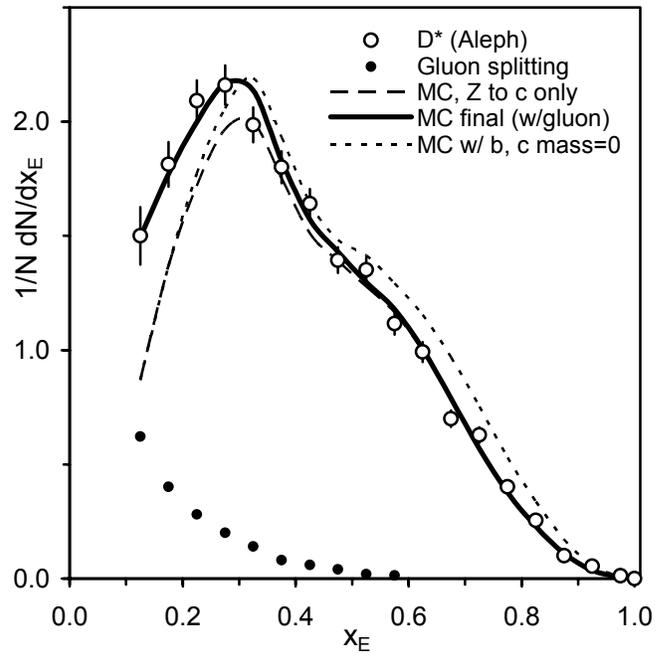

**Fig. 8**

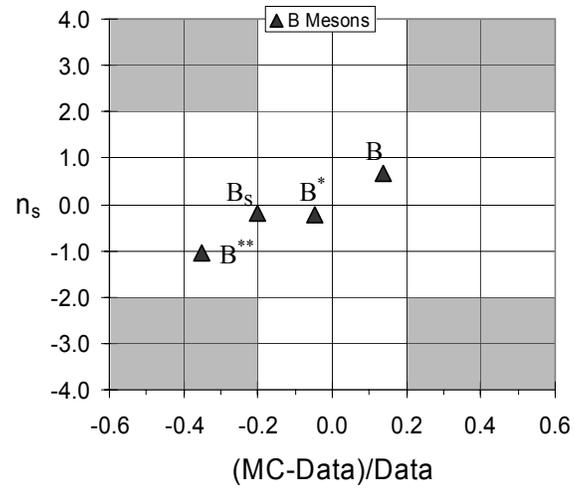

**Fig. 9**

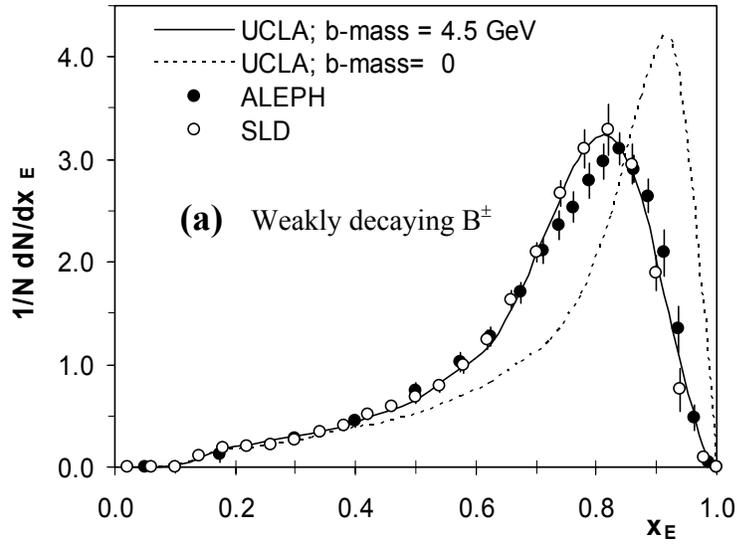

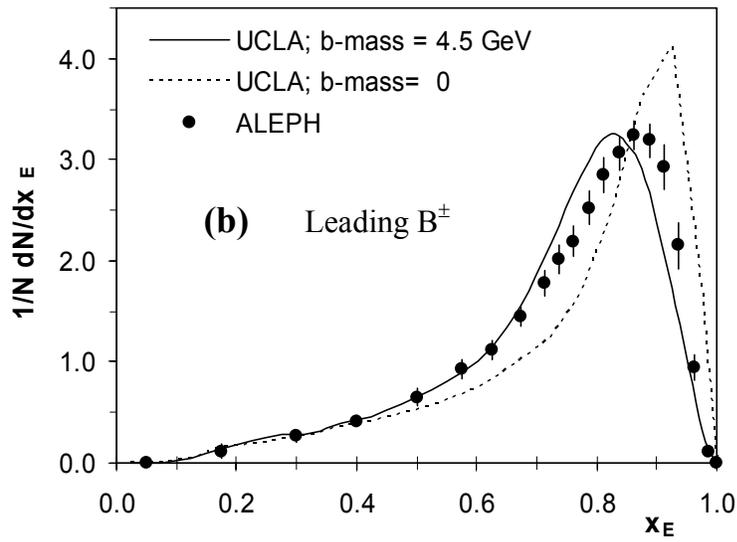

**Fig. 10**

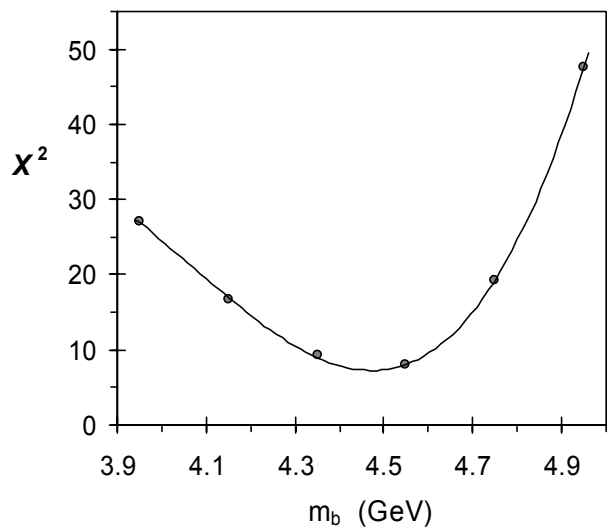

**Fig. 11**

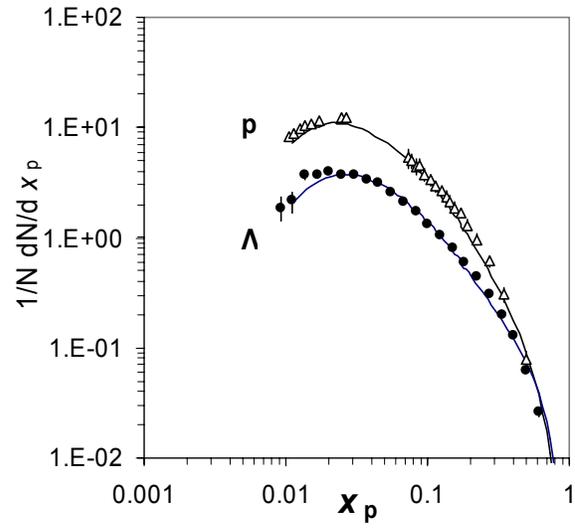

**Fig. 12**

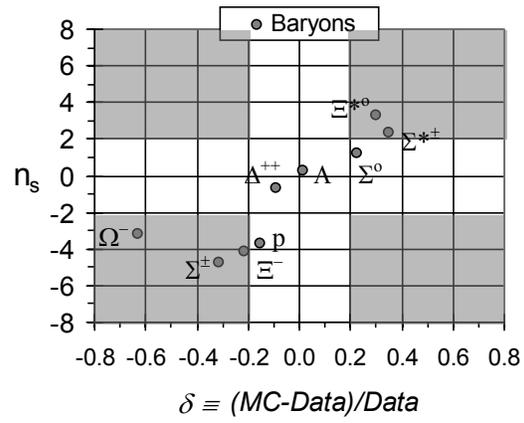

**Table 1.**

|  |  | (% probability) |  |  |
|---|---|---|---|---|
| Source | Meson | Data | UCLA | $n_s$ |
| $c \rightarrow$ | $D^{*\pm}$ | $23.2 \pm 1.5$ | 20.8 | −1.6 |
|  | $D_s^{*\pm}$ | $6.9 \pm 2.6$ | 7.9 | 0.4 |
|  | $D_{s1}^{\pm}$ | $0.94 \pm 0.23$ | 0.90 | −0.2 |
|  | $D_{s2}^{*\pm}$ | $1.14 \pm 0.60$ [†] | 0.86 | −0.5 |
| $b \rightarrow c \rightarrow$ | $D_s^{*\pm}$ | $11.3 \pm 4.6$ | 16.7 | 1.2 |
|  | $D_{s1}^{\pm}$ | $0.55 \pm 0.20$ | 0.18 | −1.9 |
|  | $D_{s2}^{*\pm}$ | $0.57 \pm 0.64$ | 0.31 | −0.4 |
| *Total hadronic* | $D_{s1}^{\pm}$ | $0.52 \pm 0.11$ | 0.34 | −1.6 |
|  | $D_{s2}^{*\pm}$ | $0.83 \pm 0.29$ | 0.51 | −1.1 |
| $\chi^2$ : | (*9 DOF*) |  |  | 12.0 |

**Table 2.**

|  |  | $m_b = 0$ GeV | | $m_b = 4.5$ GeV | |
| --- | --- | --- | --- | --- | --- |
| Meson | Data | UCLA | ($n_s$) | UCLA | ($n_s$) |
| $B^\pm$ or $B^o$ | $0.165 \pm 0.026$ | 0.201 | (1.4) | 0.187 | (0.9) |
| $B^o_s$ | $0.057 \pm .013$ | 0.032 | (–2.0) | 0.045 | (–0.9) |
| $B^*$ | $0.288 \pm 0.026$ | 0.250 | (–1.7) | 0.274 | (–0.5) |
| $B^{**}$ | $0.118 \pm .024$ | 0.0044 | (–4.7) | 0.076 | (–1.8) |
| $\chi^2$ : | (*4 DOF*) | 30.9 | | 5.1 | |